\documentclass{ecai}
\usepackage[dvipdfmx]{graphicx}
\usepackage{latexsym}
\usepackage{algorithm}
\usepackage{algorithmicx}
\usepackage[noend]{algpseudocode}
\usepackage{url}
\usepackage{amsmath}
\usepackage{amssymb}
\usepackage{amsfonts}
\usepackage{mathtools}
\usepackage{stmaryrd}
\usepackage{bm}
\usepackage{cite}
\usepackage{enumitem}
\usepackage{pifont}

\usepackage{xcolor}

\DeclareMathOperator{\Enc}{\mathsf{Enc}}
\DeclareMathOperator{\Dec}{\mathsf{Dec}}

\begin{document}

\title{
  NASS: Optimizing Secure Inference via Neural Architecture Search
}
\author{
Song Bian\institute{Kyoto University, Kyoto, Japan, email: paper@easter.kuee.kyoto-u.ac.jp}\and 
Weiwen Jiang${^2}$\and 
Qing Lu$^{2}$\and 
Yiyu Shi\institute{University of Notre Dame, Indiana, USA, email:\{wjiang2, qlu2, yshi4\}@nd.edu}\and 
Takashi Sato$^{1}$
}

\maketitle
\newcommand{\BoldVec}[1]{
  \expandafter\def\csname b#1\endcsname{{\bf{#1}}}%
}
\newcommand{\InnerProd}[2]{
  \expandafter\def\csname in#1#2\endcsname{\text{$\langle{\bf{#1}},{\bf{#2}}\rangle$}}%
}
\newcommand{\mRound}[1]{
  \text{$\left\lfloor #1\right\rceil$}%
}
\newcommand{\numberthis}{
  \addtocounter{equation}{1}\tag{\theequation}
}
\BoldVec{x}\BoldVec{y}\BoldVec{a}\BoldVec{s}\BoldVec{c}\BoldVec{r}\BoldVec{p}\BoldVec{e}\BoldVec{b}
\BoldVec{u}\BoldVec{v}\BoldVec{w}\BoldVec{q}\BoldVec{n}\BoldVec{t}
\InnerProd{a}{s}
\newcommand{\ElementMult}[2]{
  #1\boxcircle #2
}

\def\zz{{\mathbb{Z}}}
\def\zr{{\mathbb{R}}}
\def\crr{{\mathcal{R}}}
\def\crk{{\mathcal{K}}}
\def\cra{{\mathcal{A}}}
\def\Conv{{\mathsf{Conv}}}
\def\FC{{\mathsf{FC}}}
\def\NTT{{\mathsf{NTT}}}
\def\INTT{{\mathsf{INTT}}}
\def\Perm{{\mathsf{Perm}}}
\def\rot{{\mathsf{rot}}}
\def\Decomp{{\mathsf{Decomp}}}
\def\bpt{b}
\def\etarot{\eta_{\rm rot}}
\def\etamult{\eta_{\rm mult}}
\def\aaprox{\alpha}
\def\fe{{f_{\rm E}}}
\def\fd{{f_{\rm D}}}
\def\bB{{\mathcal{B}}}
\def\ee{{\varepsilon}}
\def\bee{{\bm{\varepsilon}}}
\def\bpi{{\bm{\pi}}}
\def\etal{et~al.\ }
\def\clgq{\lceil\lg{q}\rceil}
\def\clgp{\lceil\lg{p}\rceil}
\algnewcommand\algorithmicforeach{\textbf{for each}}
\algdef{S}[FOR]{ForEach}[1]{\algorithmicforeach\ #1\ \algorithmicdo}
\def\placeholder{NASS}
\def\SIMDScMult{{\mathsf{SIMDScMult}}}

\def\hatu{{\hat{u}}}
\def\hats{{\hat{s}}}
\def\hatw{{\hat{w}}}
\def\ovu{{\overline{u}}}
\def\ovw{{\overline{w}}}
\def\ovc{{\overline{c}}}
\def\ovbc{{\overline{\bc}}}
\def\hatU{{\hat{U}}}
\def\hatR{{\hat{R}}}
\def\hatW{{\hat{W}}}
\def\hatbu{{\hat{\bu}}}
\def\hatbr{{\hat{\br}}}
\def\hatbw{{\hat{\bw}}}
\def\hatbv{{\hat{\bv}}}
\def\hatbx{{\hat{\bx}}}
\def\hatby{{\hat{\by}}}
\def\hatbs{{\hat{\bs}}}
\def\bpt{b}
\def\etarot{\eta_{\rm rot}}
\def\etamult{\eta_{\rm mult}}
\def\aaprox{\alpha}
\def\fe{{f_{\rm E}}}
\def\fd{{f_{\rm D}}}
\def\ee{{\varepsilon}}
\def\bee{{\bm{\varepsilon}}}
\def\bpi{{\bm{\pi}}}
\def\etal{et~al.\ }
\def\clgq{\lceil\lg{q}\rceil}
\def\clgp{\lceil\lg{p}\rceil}
\def\pass{p_{\rm A}}
\def\share{\mathsf{Share}}
\def\rec{\mathsf{Rec}}
\def\networkparm{\mathsf{NetworkParms}}
\def\layerparm{\mathsf{LayerParms}}
\def\layerparml{\layerparm_{\rm L}}
\def\layerparmnl{\layerparm_{\rm NL}}

\begin{abstract}
Due to increasing privacy concerns, neural network (NN) based secure
inference (SI) schemes that simultaneously hide the client inputs and server
models attract major research interests. While existing works focused on
developing secure protocols for NN-based SI, in this work, we take a different
approach. We propose {\placeholder}, an integrated framework to search for
tailored NN architectures designed specifically for SI. In particular, we propose to
model cryptographic protocols as design elements with associated reward
functions.  The characterized models are then adopted in a
joint optimization with predicted hyperparameters in identifying the best NN
architectures that balance prediction accuracy and execution efficiency.  In
the experiment, it is demonstrated that we can achieve the best of both worlds
by using {\placeholder}, where the prediction accuracy can be improved from
81.6\% to 84.6\%, while the inference runtime is reduced by 2x and
communication bandwidth by 1.9x on the CIFAR-10 dataset.


\end{abstract}

\setlength{\textfloatsep}{6pt}
\setlength{\floatsep}{6pt}
\setlength{\dbltextfloatsep}{6pt}

\section{Introduction}\label{sec:intro}
With serious concerns growing over the security risks of property
stealing~\cite{juuti2019prada} and private information
leakage~\cite{shokri2017membership} related to machine learning as a service schemes, 
the study of the security properties of both neural network (NN)
training~\cite{liu2017oblivious} and inference~\cite{juvekar2018gazelle} is
becoming one of the most important fields of study across the disciplines. In
particular, a secure inference (SI) scheme refers to the situation where Bob
as a client wants to hide his inputs to Alice, the NN service provider.
Meanwhile, Alice also needs to protect her trained network model, as such a
trained model is extremely valuable due to the costly dataset preparation and
lengthy training processes.

While a number of protocols have been proposed for both secure inference and
training on neural networks~\cite{liu2017oblivious, mohassel2017secureml,
rouhani2018deepsecure, juvekar2018gazelle}, the general approach of existing
works is to find the equivalent NN operations (e.g., matrix-vector product,
activation functions) in the secure domain (e.g., using garbled circuits or
homomorphic encryption), and instantiate the secure protocols accordingly. In
other words, security is not an integral part of the proposed protocol, but
rather an added feature with (in many cases, serious) performance penalties. 

Recent advances in the secure machine learning field have suggested the
possibility of formulating the secure protocols as a design automation problem.
For example, in~\cite{bian2019darl}, authors proposed a framework that
automatically instantiate parameters for homomorphic encryption (HE) schemes.
Likewise, a line of research efforts~\cite{juvekar2018gazelle, jiang2018secure,
dathathri2019chet} have explored how to optimize HE parameters and packing
capabilities to improve the efficiency of secure computations, especially for
neural network based protocols.  In all of the existing works, secure primitives are
designed to maximize efficiency of a pre-defined neural
architecture (in fact, many of the existing works use the same manually
designed architecture).

\begin{figure}[!t]
 \centering
 \includegraphics[width=0.75\columnwidth]{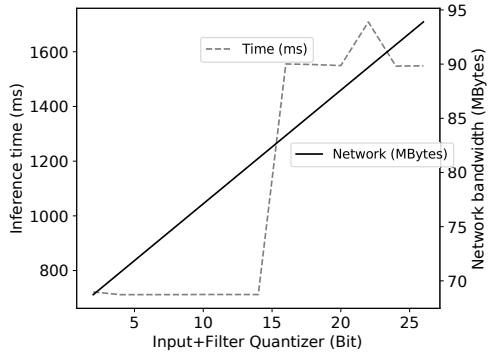}
 \caption{The relationship between neural architectures and the performance of
 secure inference, using a network with 1 Conv, 1 Relu, and 1 FC as example,
 where one more bit may double the inference time.}
 \label{fig:motivation}
\end{figure}

We argue that the existing design techniques based on fixed neural
architectures lead to unsatisfactory solutions, as the efficiency of SI (in
terms of the inference time and network bandwidth) are significantly affected
by the architectures.  The performance non-linearity of cryptographic
primitives are demonstrated through Fig.~\ref{fig:motivation}, where the
computational and communication costs of SI are plotted with respect to
quantization factors of some neural architecture. From
Fig.~\ref{fig:motivation}, we can see that for certain quantization intervals
(e.g., 14-bit to 15-bit), the inference time is doubled, while for other
intervals (e.g., from 2-bit up to 15 bits), the inference time remains
unchanged. This non-linear performance curve is primarily due to the
underlying primitive (in this case a packed homomorphic encryption (PAHE)
scheme) that is constrained by its cryptographic parameters.
Our conclusion here is
that, in order to obtain better design trade-offs, a joint exploration
considering both secure primitives and neural architectures is required to push
forward the Pareto frontiers of the efficiency and prediction accuracy of
NN-based SI.

In this work, we propose {\placeholder}, a novel Neural Architecture Search
framework for Secure inference, where the optimization of cryptographic
primitives and NN prediction accuracy are integrated. To the best of our 
knowledge, we are the first to take a
{\it{synthetic}} approach to improve both the accuracy and the efficiency of
SI on NN. In {\placeholder},  the process of finding the best SI scheme is formulated 
as predicting the most rewarding neural architecture, where
the reward is derived from the
accuracy and efficiency statistics. A 
system optimizer based on reinforcement learning is used to
take feedback from the rewards to generate new architectures, acting as a neural
architecture search (NAS) engine. 
Our main contributions are summarized as follows.


\begin{itemize}
 \setlength{\itemsep}{0pt}
  \item {\bf{Synthesizing Secure Architectures}}:
    To the best of our knowledge, {\placeholder} represents the first work to
    search for neural architectures optimized in secure applications with
    multiple cryptographic building blocks.  In {\placeholder},
    cryptographic primitives are modeled as design elements, and secure
    computations with these elements become abstract operators that can be
    automatically synthesized by the optimization engines. 
    
  \item {\bf{Optimizing HE Parameters}}:
    While existing works have already treated the instantiation of HE
    parameters as a design problem and proposed some
    solutions~\cite{bian2019darl}, we point out that these solutions
    are not adequate. In particular,
    we identify an optimization dilemma in learning with errors (LWE) based HE
    parameter instantiation, and observe that this optimization problem is
    (computationally) rather difficult to solve, especially for NAS-based
    optimization with fast turnaround time.

  \item {\bf{A Thorough Architectural Search for SI}}:
    By conducting extensive architectural search, it is demonstrated that the
    performance of SI can be reduced while improving the prediction accuracy.
    We achieve a prediction accuracy of 84.6\% on the CIFAR-10 dataset, while
    reducing 2x computational time and 1.9x network bandwidth, compared to
    the best known SI scheme~\cite{juvekar2018gazelle} with a prediction accuracy
    of only 81.6\%.
\end{itemize}

The rest of this paper is organized as follows. First, in
Section~\ref{sec:background}, basics on PAHE, secure inference, and NAS are
discussed.  Second, the {\placeholder} framework is outlined in
Sections~\ref{sec:framework} and~\ref{sec:propose}, where we detail how
security and parameter analyses can be systematically performed, along with
the design of reward functions for the integration with the NAS engine. Next,
the output architectures of {\placeholder} along with performance statistics
are demonstrated in Section~\ref{sec:experiment}.  Finally, our work is
summarized in Section~\ref{sec:conclusion}.

\section{Preliminaries}\label{sec:background}

\subsection{Cryptographic Building Blocks}\label{sec:pahe}
In this work, we mainly consider the optimization involving two types of
cryptographic primitives, packed additive homomorphic encryption (PAHE) based
on the ring learning with error (RLWE) 
problem~\cite{brakerski2012fully, fan2012somewhat, brakerski2012leveled, cheon2017homomorphic}, 
and garbled circuits (GC)~\cite{yao1982protocols}. In
what follows, we provide a high-level abstraction of each individual primitive.

{\bf{PAHE}}: A PAHE is a cryptosystem, where the encryption ($\Enc$) and
decryption ($\Dec$) functions act as group (additive) homomorphisms between
the plaintext and ciphertext spaces. Except for the normal $\Enc$ and
$\Dec$, a PAHE scheme is equipped with the following three
abstract operators. We use $[\bx]$ to denote the encrypted ciphertext of
$\bx\in\zz^{n}$, and $n\in\zz$ here is some lattice dimension.
\begin{itemize}
  \item Homomorphic addition $(\boxplus)$: for $\bx, \by\in\zz^{n}$, 
    $\Dec([\bx]\boxplus[\by])=\bx+\by$.
  \item Homomorphic Hadamard product $(\boxcircle)$: for $\bx, \by\in\zz^{n}$,
    $\Dec(\ElementMult{[\bx]}{\by})=\bx\circ \by$, where $\circ$ is the
    element-wise multiplication operator.
  \item Homomorphic rotation $(\rot)$: for $\bx\in\zz^{n}$, let
  $\bx=(x_0, x_1, \cdots, x_{n-1})$, 
    $\rot([\bx], k)=(x_{k}, x_{k+1}, \cdots, x_{n-1}, x_{0}, \cdots, x_{k-1})$
  for $k\in\{0, \cdots, n-1\}$.
\end{itemize}

{\bf{GC}}: GC can be considered as a more general form of HE. In particular,
the circuit garbler, Alice, ``encrypts'' some function $f$ along with her input
$x$ to Bob, the circuit evaluator. Bob evaluates $f(x, y)$ using his encrypted
input $y$ that is received from Alice obliviously, and obtains the encrypted
outputs. Alice and Bob jointly ``decrypt'' the output of the function $f(x,
y)$ and one of the two parties learn the result.

\subsection{Homomorphic Evaluation Errors in RLWE-based PAHE}\label{sec:error_size}
In this work, we omit details on the implementation of RLWE-based PAHE schemes,
such as BFV~\cite{brakerski2012fully, fan2012somewhat},
BGV~\cite{brakerski2012leveled}, and CKKS~\cite{cheon2017homomorphic}.
However, for all of the above RLWE-based PAHE schemes (and most RLWE-based
cryptosystems), the ciphertext output from the encryption function of the
cryptosystem bares some intrinsic errors, which can be thought of an
additive components to the ciphertext, i.e., 
\begin{align}
  \ovbc = \bc + \be,
\end{align}
where $\bc$ is the ``errorless'' ciphertext, and $\be$ the error (both are
vectors in $\zz^{n}$ for some lattice dimension $n$). It is obvious that when
we add two ciphertexts, $\ovbc_{0}\boxplus\ovbc_{1}$, the error is also
additively increased (i.e., $\be_{\rm{sum}}=\be_{0}+\be_{1}$). Similarly,
homomorphic Hadamard product and rotation operations also increases the
errors. Therefore, each level of
homomorphic evaluation increases the error contained in the ciphertext, and
when the size of the error become too large (i.e., too many levels of homomorphic
evaluations), {\it{some}} ciphertexts will not be correctly deciphered.  We
emphasize the point that not all ciphertexts become undecipherable as the size
of the error is randomly distributed, and this probabilistic behavior can be
utilized to improve the efficiency of SI schemes.


\subsection{Secure Neural Network Inference}\label{sec:si}
\begin{figure}[!t]
 \centering
 \includegraphics[width=0.99\columnwidth]{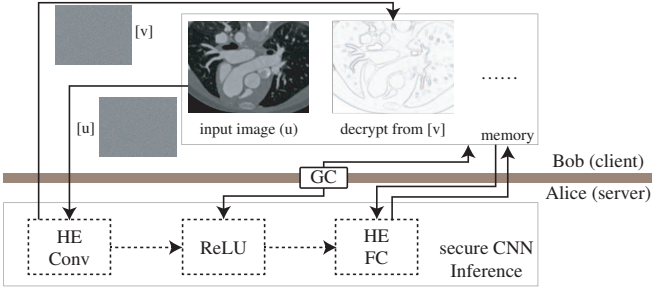}
 \caption{An example of the architecture in Gazelle with one $\Conv$ layers,
   one non-linear layers and one $\FC$ layer.}
 \label{fig:gazelle_gen}
\end{figure}
While a number of pioneer works have already established the concept of secure
inference and training with neural networks~\cite{liu2017oblivious,
mohassel2017secureml, rouhani2018deepsecure}, it was not until recently that
such protocols carried practical significance. For example,
in~\cite{liu2017oblivious}, an inference with a single CIFAR-10 image takes
more than 500 seconds to complete. Using the same neural architecture, the
performance was improved to less than 13 seconds in one of the most recent
arts on SI, Gazelle~\cite{juvekar2018gazelle}. Unfortunately, 13 seconds per
image inference is obviously still unsatisfactory, especially given the large
amount of data exchange in real-world applications. Therefore, we adopt the
Gazelle protocol in this work, and take a system-level approach to improve its
efficiency.

An overview of the Gazelle protocol is outlined in
Fig.~\ref{fig:gazelle_gen}, where Alice wants to classify some input (e.g.,
image), and Bob holds the weights. The Gazelle protocol classifies all NN
operations into two types of layers: i) linear layers, where the computations
are efficiently carried out by PAHE-based cryptographic primitives, and ii)
non-linear layers, where interactive protocols such as multiplication
triples~\cite{beaver1991efficient} or GC
are employed.

{\bf{Threat Model}}: The threat model in Gazelle and this work is that both Bob
and Alice are semi-honest, in the sense that both parties follow the described
protocol (e.g., encryption and decryption procedures in PAHE, GC), but want to
learn as much information as possible from the other party. In particular,
Alice wishes to gain knowledge on the trained model from Bob, and Bob is
curious about the encrypted inputs from Alice.



\subsection{Neural Architecture Search}

Recently, Neural Architecture Search (NAS) has been consistently breaking
the accuracy records in a variety of machine learning applications, such as image
classification~\cite{zoph2016neural}, image segmentation~\cite{liu2019auto},
video action recognition~\cite{peng2019video}, and many more. NAS attracts major
attentions mainly because it successfully eliminates the needs of human
expertise and labor time in identifying high-accuracy neural architectures.

A typical NAS, such as that in \cite{zoph2016neural}, is composed of a
controller and a trainer.  The controller will iteratively predict (i.e.,
generate) neural architecture parameters, referred to as child networks. The
child networks will be trained from scratch by the trainer on a held-out dataset to
obtain the prediction accuracy.  Then, the accuracy will be feedback to update the
controller.  Finally, after the number of child networks predicted by the
controller exceed a predefined threshold, the search process will be
terminated.  The searched architecture with the highest accuracy is identified 
to be the output of the NAS engine.

Existing works have demonstrated that the automatically searched neural
architectures can achieve close accuracy to the best human-invented
architectures~\cite{zoph2017learning,zoph2016neural}. In addition,
we also identify multi-objective NAS techniques proposed under the context
of field-programmable gate array (FPGA) and mobile 
platforms~\cite{yang2018optimal,elsken2018efficient,jiang2019hardware,
jiang2019device,lu2019neural,zhang2019neural,tan2019mnasnet,yang2020coexplore}.
However, without proper security performance measures, the identified 
architectures can have over-complex 
structures that render them useless in real-world cryptographic applications.
In addition, as demonstrated in Fig.~\ref{fig:motivation}, cryptographic primitives 
generally have complicated performance trade-offs, and
no existing works have demonstrated that
a multi-objective NAS engine is able to learn such complex behaviors.
Therefore, the main motivation
of the {\placeholder} framework is to find accurate and efficient 
neural architectures for secure inference schemes.

\section{\placeholder~Framework}\label{sec:framework}

\subsection{Problem Formulation and Challenges}

In this paper, we aim to identify the most efficient secure neural network
inference via neural architecture search.
The problem is informally defined as
follows: Given a specific dataset and a set of secure inference protocols, our
objective is to automatically generate a quantized neural network architecture 
and the parameters for each of the cryptographic primitives,
such that the reward of the resultant neural network after training can be maximized. Here, we
define the reward to be a function of the prediction accuracy and performance statistics, including
the inference time and network bandwidth.

To solve the above problem, several challenges need to be addressed from both
the neural architecture search perspective and the secure protocol perspective.
We list two main challenges as follows. 

\textit{Challenge 1:} There are missing links among neural architectural optimizations, quantization optimizations, and cryptographic protocol optimizations, resulting in the non-optimal solutions from
existing works.
This is based on our observation that all of the above optimizations are tightly cross-coupled; that is, 
the optimization in one direction (e.g., better prediction accuracy) can have positive or negative impact on the other directions (e.g., larger quantization level and higher secure inference time).
Therefore, a framework that can jointly optimize neural architectures, quantizations, and performances of cryptographic primitives, is needed.
In this work, we derive the {\placeholder}~framework
(Section~\ref{Sec:Framework} and~\ref{Sec:FrameworkDetails}) to fill the gap.

\textit{Challenge 2:} To the best of our knowledge, there exists no efficient
performance estimator for secure inference (SI) involving multiple cryptographic primitives. Since the NAS engine generates a large amount of intermediate architectures iteratively,
without an automatic performance estimator, it is impossible to evaluate the performance statistics of such networks adopted in SI. 
In this paper, we make the contribution of developing efficient estimator engines (Section
\ref{sec:propose}).

\begin{figure}[!t]
 \centering
 \includegraphics[width=1\columnwidth]{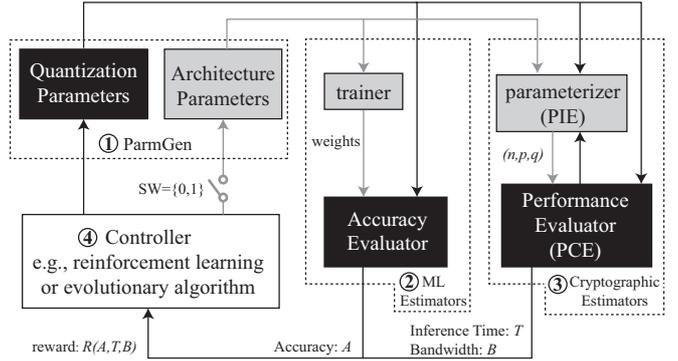}
 \vspace{-20pt}
 \caption{An overview on the proposed {\placeholder}~framework.}
 \label{fig:NASS}
\end{figure}

\subsection{Overview of {\placeholder}}\label{Sec:Framework}

Fig. \ref{fig:NASS} illustrates an overview of the proposed
{\placeholder}~Framework with four components:
{\large\ding{192}}
{ParmGen}, {\large\ding{193}} {Machine Learning (ML) Estimators},
{\large\ding{194}} {Cryptographic Estimators}, and {\large\ding{195}}
{Controller}. 
Specifically, component {\large\ding{192}} parameterizes the architecture and quantization, 
which identifies a unique neural architecture for the subsequent computations. Upon receiving
the input architecture from component {\large\ding{192}}, Component {\large\ding{193}} 
trains and evaluates its accuracy, and component {\large\ding{194}} 
optimizes the  cryptographic primitives to estimates its performance. 
Finally, component {\large\ding{195}} will 
control the optimization flow. All of the components collaboratively explore the parameter spaces of 
neural architecture, quantization, and cryptographic primitives to jointly optimize the accuracy, 
time, and bandwidth.

The {\placeholder}~framework works in three steps. First, the controller generates 
a prediction on a quantized neural architecture (called child
network), which will be formulated as $\layerparm$.  
Second, the child network will be evaluated by ML Estimators to generate prediction accuracy
(A), and optimized in Cryptographic Estimators to provide the inference time (T), and bandwidth
(B) feedback.
Lastly, a reward signal is generated in terms of A, T, B, to update the controller.
Details of each component will be introduced in Section
\ref{Sec:FrameworkDetails}.

In practice, the lengthy training process dominates the search time.
In {\placeholder}, we add a switch $SW$ before the architecture parameter (AP) subcomponent to 
dramatically reduce the number of training processes. 
This is based on the observation that the quality of quantization parameters for the same
architectures can be evaluated using the same trained (floating-point) weights.
If the switch is on ($SW=1$), we will train the architecture from
scratch to generate weights and obtain accuracy statistics in terms of quantization.
Otherwise ($SW=0$), we reuse the weights and apply the new quantization parameters to 
obtain the accuracy.

The switch can be controlled by using a predefined function.  In this work, we
demonstrate the exploration procedure using the following function:
\begin{equation}\label{equ:reward}
SW = \left\{ {\begin{array}{*{20}{c}}
{1}&{Eps\ mod\ SW_N = 0}\\
{0}&{Otherwise}
\end{array}} \right.
\end{equation}
where $Eps$ is the episode index given to each of the child networks predicted by the controller,
$SW_N$ is a scalar to indicate the number of child network with the
same architecture but different quantization to be explored.






\subsection{\placeholder~Framework Details}\label{Sec:FrameworkDetails}

{\large\ding{192}} \textbf{{ParmGen.}}
The ParmGen block generates layer parameters for the subsequent computations. 
A neural architecture consists of a set of layers.  According to the linearity
of function for each layer, there are two types of layers: linear layer (e.g.,
convolution, fully connection), and non-linear layer (e.g., ReLU,
pooling).  Each layer can be specified as a set of parameters.  Note
that different types of layers have different parameters.  We denote
$\layerparml$ and $\layerparmnl$ to represent parameters of linear layers and
non-linear layers, respectively.

For \textit{linear layers}, a set of layer parameter is denoted as
$\layerparml=\{n_{i}, n_{o}, f_{w}, f_{h}, l_{i}, l_{f}, c_{i}, c_{o}\}$,
where $n_{i}$ and $n_{o}$ represent the dimensions of feature maps (i.e.,
input data); $f_{w}$ and $f_{h}$ represent the dimensions of filters (i.e.,
weights), $l_{i}$ and $l_{f}$ indicate the data and weight quantizers; $c_{i}$
and $c_{o}$ stand for the number of input and output channels.

For \textit{non-linear layers}, they do not contain weights, and therefore,
there is no parameters for filter quantizations and dimensions. In addition, the number
of channels will not be changed, and we only record the input channel number
$c_{i}$ in layer parameter $\layerparmnl$.  In consequence, we denote
parameter sets for non-linear layers as $\layerparmnl=\{n_{i}, n_{o}, l_{i},
l_{o}, c_{i}\}$.

A neural architecture can be represented by a collection of parameters for all
layers, i.e., $\networkparm=\{\layerparm_{{\rm L}, k}, \layerparm_{{\rm NL}, k}\}$ where
$k=0, 1, \cdots$ represents the $k^{th}$ layer.

{\large\ding{193}} \textbf{{Machine Learning (ML) Estimator.}}
A machine learning estimator is composed of a trainer and an accuracy
evaluator.  According to the status of switch before AP in {\large{\ding{192}}},
the ML estimator will take different actions.  When
the switch is on (i.e., $SW=1$), a new architecture will flow into the ML
estimator, and it will be trained from scratch using floating points. Then,
the accuracy evaluator will quantize the weights from the trainer according to the
given quantization parameters to obtain the accuracy of the quantized neural
network.  When the switch is off (i.e., $SW=0$), it indicates that the
previous predicted architecture is applied with new quantization parameters.
In this case, the ML estimator will not train the architecture. 
The weights from in the previous iteration are reused with new quantization
parameters to obtain the prediction accuracy on the training dataset.


{\large\ding{194}} \textbf{{Cryptographic Estimator.}} A cryptographic estimator
contains two sub-components: the parameter instantiation engine (PIE) 
and the performance characterization engine (PCE).  These engines take input from
$\layerparm$, and collaboratively instantiate parameters for the cryptographic
parameters while evaluating their performance.
In particular, the outputs of PIE are the cryptographic parameters.  For
example, for
RLWE-based PAHE schemes (e.g., used for homomorphic matrix-vector product in linear layers and
multiplication triples for square activation), the cryptographic parameters
are $(n, q, p)$.
During performance characterization, PCE 
consults with PAHE and GC libraries to produce characterized scores for
a single round of secure inference using the architecture specified in
$\networkparm$.  Details on the implementations of PIE and PCE will be
discussed in Section~\ref{sec:propose}. Kindly note that, for some 
cryptographic protocols (e.g., GC used in implementing ReLU), 
the cryptographic parameters can be directly determined in terms of
$\layerparmnl$ without using a PIE.



{\large\ding{195}} \textbf{{Controller.}}
The controller is a core component in the {\placeholder} framework. According to
the output of the ML estimator ({\large{\ding{193}}}) and the cryptographic estimator 
({\large{\ding{194}}}), the controller predicts a new $\networkparm$ which supposedly has 
higher accuracy, lower latency, and lower bandwidth requirement compared to the 
architecture predicted in the previous iteration.

The controller can be implemented by different techniques, such as the
reinforcement learning or the evolutionary algorithms.  However, in both
cases, the key element for the controller design is the reward function.  In
this work, we employ the reinforcement learning method in the controller whose
interactions with the environment are modeled as a Markov decision process
(MDP).  The reward function is formulated as follows.
\begin{align}\label{eq:rewarding}
  R(A, T, B) = A+A\cdot \xi(T, B),
\end{align}
where A is the prediction accuracy, and $\xi(T, B)$ is the performance
score reported by the cryptographic estimators. The detailed definition of
$\xi$ can be found in Eq.~\eqref{eq:performance_score}.  After calculating the
reward, we follow the Monte Carlo policy gradient algorithm
\cite{WILLIAMS1992Simple} to update the controller:
\begin{equation}
    \triangledown J(\theta) = \frac{1}{m}\sum\limits_{k=1}^m\sum\limits_{\tau=1}^t\gamma^{t-\tau}\triangledown_\theta\log \pi_\theta(a_\tau|a_{\left(\tau-1\right):1})(R_k-b)
\end{equation}
where $m$ is the batch size and $t$ is the total number of steps in each
episode. The rewards are discounted at every step by an exponential factor
$\gamma$ and the baseline $b$ is the exponential moving average of the rewards.

\section{Estimators for Cryptographic Primitives}\label{sec:propose}
While CHET~\cite{dathathri2019chet} realizes the importance of establishing an
abstraction layer for the NN designer to hide specific HE implementation
details, they did not think of the cryptographic primitives as {\it{design
elements}} that carry distinct performance trade-offs (actually, CHET only
focused on compiling a single FHE primitive). As observed in Fig.~\ref{fig:gazelle_gen}, 
Gazelle instantiate different protocols according to the specific NN layers.
Hence, in this section, we describe how to
construct estimators that model cryptographic primitives as delay elements
with communicational costs, analogous to the FPGA components modeled in the
FNAS~\cite{jiang2019accuracy} framework.

\subsection{Constructing PIE for PAHE}\label{sec:feedback}
In this work, we use the widely-adopted BFV~\cite{fan2012somewhat} cryptosystem
as the example PAHE scheme, but our method applies broadly to all RLWE-based
PAHE cryptosystems.  In BFV, three parameters are required to instantiate the
cryptosystem, $(n, p, q)$, where $n$ is the lattice dimension, $p$ the
plaintext modulus, and $q$ the ciphertext modulus. 

\subsubsection{The Feedback Loop}
In the Gazelle protocol, since each linear layer is evaluated independently
(decryptions are performed after only one layer of homomorphic evaluation),
parameters can be minimized. For example, in our experiments, $n=(2048,
4096, 8192)$, and $q$ ranges from 60 to 180 bits. Therefore, even one bit of
loose error margin can easily result in 1.5x to 2x performance penalty on
64-bit machines, due to the requirement of extra integer slots (e.g., from
61-bit $q$ to 62-bit $q$). 

In Gazelle, as long as the dimensions and quantizers are the same, the
parameters do not scale with the number of layers in the NN.  Therefore,
parameter minimization needs to be carried out for every NN layer with varying
quantization and filter dimensions. The main difficulty for per-layer parameter minimization lies
in the feedback loop between PCE and PIE. The dilemma is that, in order for
PIE to instantiate parameters that ensure correct decryption, the error size
(explained in Section~\ref{sec:error_size})
needs to be estimated by PCE.  Meanwhile, PCE needs instantiated PAHE
parameters from PIE to perform error analysis, thereby forms the loop.
Iterating through all possible parameter combinations with error calculations
for each NN layer creates significant computational burden in the
{\placeholder} optimization process. In addition, generating large primes can 
also be time consuming, as BFV requires additional
constraints on the relationship of $p$, $q$ and $n$ to enable the batching
technique~\cite{smart2010fully}, which is essential to the efficiency of SI.
In particular, both $p$ and $q$ need to satisfy $p\equiv q\equiv 1
(\bmod{\hspace{0.3em}n})$,  where $q$ can be a large integer (e.g., 120 bits).

\subsubsection{Instantiating the Parameters}

\begin{figure}[!t]
 \centering
 \includegraphics[width=1\columnwidth]{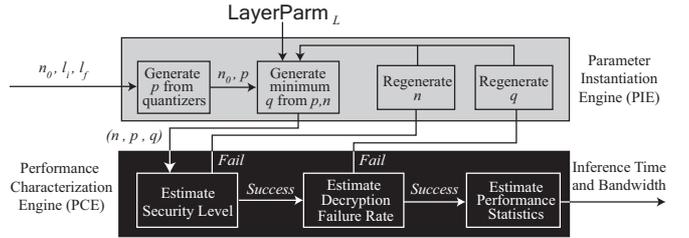}
 \vspace{-20pt}
 \caption{The PIE-PCE co-optimization procedures for characterizing 
 the performance cost of a linear layer.}
 \label{fig:pie_pce}
\end{figure}

An overview of the joint parameter optimization procedure is illustrated in
Fig.~\ref{fig:pie_pce}.

{\large\ding{192}} {\bf{Initialization}}: To start the optimization process,
inputs are first fed to PIE. The inputs include $n_{0}$, the initial lattice
dimension, and ($l_{i}$, $l_{f}$), the respect quantizers for NN inputs and
filters. Here, $n_{0}$ is an arbitrary number, and can be set as the smallest
$n$ that grants some security levels for extremely small $q$ (e.g.,
$n_{0}=1024$, which is secure for $q\leq 2^{32}$). $l_{i}$ and $l_{f}$ is used to
determine the plaintext modulus $p$. In order to carry out a successful
inference, we need that $p\geq l_{i}+l_{f}+\lceil\log_{2}(f_{h}\cdot
f_{w})\rceil$ and $p\equiv 1(\bmod{\hspace{0.3em}n_{0}})$. After generating the
plaintext modulus $p$, along with $n_{0}$ and other parameters in
$\layerparml$ (e.g., the input dimensions $n_{i}, n_{o}$ and the filter
dimension $f_{h}$ and $f_{w}$), we can calculate a working ciphertext modulus $q$. Note that
this estimation can be loose, but will be tightened in PCE through optimization
iterations.

{\large\ding{193}} {\bf{Optimization Loop}}: Upon receiving parameters from PIE, PCE performs two
important evaluations: i) security level estimation, and ii) decryption failure rate 
estimation. Failure in meeting either of the conditions results in an immediate 
rejection. First, in i), The security levels are consulted with the community standard established 
in~\cite{HomomorphicEncryptionSecurityStandard}. When the security 
standard is not met, we regenerate the lattice dimension $n$ and retry the security analysis. 
Next, in ii), after obtaining a valid $n$ for the estimated $q$, a set of ciphertexts are created 
to see if $q$ is large enough for correct decryption. 
If the decryption failure rate is too high, we regenerate a larger $q$ and re-evaluate 
the security of $n$ with respect to the new $q$. After deriving valid $(n, p, q)$ that
passes all the tests, the parameters are fed into a PAHE library to characterize the
estimated amount of time and memory consumed by a single layer to calculation.

{\large\ding{194}} {\bf{Output Statistics}}: Steps {\large\ding{192}} and {\large\ding{193}} 
described above will be repeated for every 
layer in the input neural architecture, and all performance statistics are summed up
to produce a final score to be used by the overall NASS framework in searching for a 
better neural architecture for SI.

\subsubsection{Generating a Valid Ciphertext Modulus}\label{sec:ciph_modulus}
One last note on the ciphertext modulus $q$ is that,  as mentioned in 
Section~\ref{sec:pahe}, not all ciphertexts become undecipherable when $q$ is small. 
The probability that a ciphertext becomes undecipherable is called the decryption
failure rate. Observe that different from~\cite{bian2019darl}, we do not need an 
expensive simulation to ensure an asymptotically small (e.g., $2^{-40}$) decryption
failure probability, since NN-based SI mispredicts much more often than $2^{-40}$.
In most cases, a 0.1\% accuracy degradation is not noticeable for practical CNN 
applications. Therefore, we can use the standard Monte-Carlo simulation technique to 
ensure that  $q$ is large enough to ensure that $\Pr[\Enc(\Dec(m))\not=m]<\delta$, 
where $\delta$ ranges from $10^{-3}$ (1 decryption failure in 1000 inferences) to 
$10^{-2}$  (1 in 100), depending on the prediction accuracy requirement.

\subsection{PCE: Performance Characterization}
\subsubsection{Characterizing Linear Layers}\label{sec:pce_lin}
The main arithmetic computations in both $\Conv$ and $\FC$ involve a set of
inner products between some plaintext matrix and ciphertext matrix 
(flatten as vectors) homomorphically. To compute any homomorphic inner product,
the pioneering work in~\cite{juvekar2018gazelle} proposes to align the weight
matrix with the rotating input ciphertext vector to minimize the number of
homomorphic operations.  In general, the algorithm computes the inner product
between $W\in\zz^{n_o\times n_i}_{p}$, a weight matrix, and
$[\bu]\in\zz^{n_i}_{q}$,
the encrypted input vector as follows.
\begin{align}
  [\bt] &= \sum_{i=0}^{n_o-1}\bw_i\boxdot \rot([\bu], i)\label{eq:gaz_eval}\\
        &=\bw_0\boxdot [\bu] + \cdots + \bw_{n_o -1}\boxdot \rot([\bu],
  n_o-1),\label{eq:hybrid_1}\\
  [\bv] &= \sum_{i=1}^{\lg{(n/n_o)}}\rot\left([\bt],
  \frac{n}{2^i}\right),\label{eq:hybrid_2}
\end{align}
where $[\bv]$ holds the result vector $\bv=W\bu\in\zz^{n_o}_{p}$, $\bw_{i}$'s
are the diagonally aligned columns of $W$ with dimension
$\bw_{i}\in\zz^{n}_{p}$, and $\lg{(\cdot)}$ denotes $\log_{2}{(\cdot)}$.  In
Eq.~\eqref{eq:hybrid_1}, we first rotate $[\bu]$
$n_o$ times, each time multiplying it with the aligned vectors
$\bw_i\in\{\bw_{0}, \cdots, \bw_{n_o-1}\}$. Each multiplication generates an
intermediate ciphertext that holds only {\it{one}} entry in $\bv_{i}$ with respect to
$\bw_{i}$.  Summing these ciphertexts gives us a single ciphertext that
is packed with $n/n_o$ partial sums in the corresponding inner products, 
and packed results can be summed up to obtain the final product [\bv].

It is noted that the performance non-linearity illustrated in
Fig.~\ref{fig:motivation} lies critically in the way homomorphic inner
products are computed. Take a toy example where $n_{o}\times n_{i}=10\times
1024$ and $n=1024$. The input vector $\bu\in\zz^{n_{o}}$ can be tightly stored
into a single ciphertext $[\bu]\in\zz^{n}$. Using the Gazelle algorithm, we
rotate the input ciphertext $n_{o}=10$ times, and compute 10 homomorphic
Hadamard products. However, suppose that the input dimension is somehow
$10\times 1025$. Since the lattice dimension $n$ can only be a power of 2, the
ciphertext size becomes $\zz^{2048}$ in order to hold an input vector
$\bu\in\zz^{1025}$. All subsequent homomorphic evaluations require double the
amount of computations and bandwidths compared to $n=1024$. If the amount of
accuracy improvement from $n_i=1024$ to $n_i=1025$ is marginal (e.g., $\leq
0.01\%$), then this improvement suggestion should be rejected.

The above example represents the precise procedure performed in PCE, where
all the information contained in $\layerparml$ jointly determines how packing
can be performed to maximize the protocol efficiency. The output of this
procedure is the inference time $T$ and network bandwidth $B$. 
We use a simple
weighted sum to derive the performance score $\xi$, where
\begin{align}\label{eq:performance_score}
  \xi(T, B)=\beta\cdot T + (1-\beta)\cdot B.
\end{align}
One important implementation detail is that, instead of
performing the entire calculation using an actual PAHE implementation, only basic
operations ($\boxplus$, $\boxcircle$ and $\rot$) need to be characterized. The
actual runtime and bandwidth usage can be scaled from the combinations of the
basic operations. In fact, this can be a critical performance improvement, as
secure inference is still quite slow for deep neural networks (10 to 100
seconds), and running PCE for full network characterizations can be a performance
bottleneck in the optimization process.

\subsubsection{PCE for Non-Linear Layers}
Running PCE for interactive protocols such as multiplication triples~\cite{beaver1991efficient}
and GC~\cite{yao1982protocols} is much simpler than linear layers, as these non-linear functions 
(e.g., square and ReLU) are performed on a per-element basis with fixed
functionality. The performance statistics can be characterized once, and used throughout all
layers when properly scaled.

\section{Numerical Experiments and Parameter Instantiations}\label{sec:experiment}

\subsection{Experiment Setup}
In this work, we compare {\placeholder} with the performance statistics of 
two best-performing recent works on secure inference, namely, 
Gazelle~\cite{juvekar2018gazelle} and XONN~\cite{riazi2019xonn}. 
First, we point out that the reported statistics 
are not entirely reliable in Gazelle. For example, the architecture used for a single 
CIFAR-10 inference needs 
more than 120,000 ReLU calls, and Gazelle reports 551\,ms of runtime per 10,000
ReLU evaluations. Nevertheless, the total online inference time is less than three seconds.
Since the main focus of {\placeholder} is to improve the architectural design of 
NNs, the performance of both Gazelle and the derived architectures in this work are 
characterized by the proposed performance estimator. We base our experiments on three
datasets, MNIST~\cite{lecun2010mnist},  fashion-MNIST~\cite{xiao2017online}, 
and CIFAR-10~\cite{krizhevsky2009learning}. The characterization 
experiments  are conducted with a Intel i3-6100 3.7\,Ghz 
CPU, and the architectural search is peformed using a NVIDIA P100 GPU. 
The adopted PAHE library is SEAL  version 3.3.0~\cite{sealcrypto}, and 
GC protocols are implemented using  ABY~\cite{demmler2015aby}.


\subsection{Architectural Optimization and the Pareto Frontier}

\begin{figure}[!t]
 \centering
 \includegraphics[width=0.75\columnwidth]{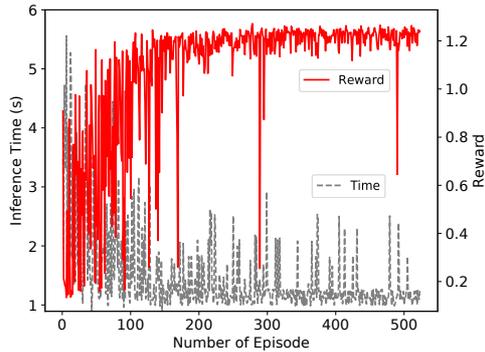}
 \caption{The learning curves show that as the number of episodes increase, 
 both the reward and the inference time tend to converge.}
 \label{fig:converge}
\end{figure}

\begin{figure}[!t]
 \centering
 \includegraphics[width=2.7361 in]{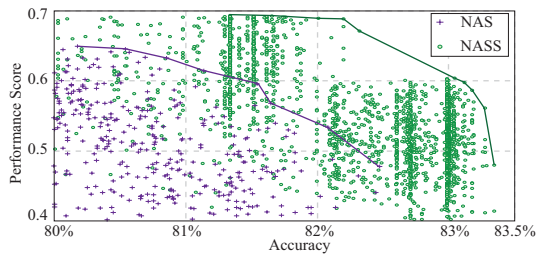}
 \caption{The proposed \placeholder~can significantly push forward Pareto frontier in terms of accuracy and score, compared with the NAS without considering secure inference: each point indicates an architecture with specific quantization parameters.}
 \label{fig:frontier}
\end{figure}

First, in Fig.~\ref{fig:converge}, we use an example {\placeholder} run using the CIFAR-10
dataset to show the effectiveness of the proposed framework. We trained the 
{\placeholder} controller through a 500 episodes window, where the
neural architecture is fixed to have four convolution layers. As 
explained in Section~\ref{Sec:Framework}, each episode generates a child network, and
the reward of the child network is calculated in Eq.~\eqref{eq:rewarding} using the 
prediction accuracy and  cryptographic performance scores. The results in Fig.~\ref{fig:converge}
indicate that both the rewards (where accuracy dominates the calculation, as described in 
Eq.~\eqref{eq:rewarding}) and the secure inference times converge to their optimized states as 
learning episodes proceed. Furthermore, in Fig.~\ref{fig:frontier}, the best-performing data 
points are gathered to  plot the Pareto frontiers generated by {\placeholder}.
Here, the vertical axis is the estimated performance score, 
and the horizontal axis denotes the
prediction accuracy on the CIFAR-10 dataset.
Two observations are made here. First, the proposed {\placeholder}
engine is able to learn the extremely non-linear design space of CNN-based SI, and second, 
the {\placeholder} framework 
pushes forward the Pareto frontier for SI compared to existing works on both SI and NAS.

\begin{table}
  \scriptsize
  \centering
  \renewcommand\arraystretch{1}
  \caption{Selected Architectures from {\placeholder}}
  \label{tab:nass_performance}
  \begin{tabular}{c|c|c|c|c}
  \hline
    Architecture & Accuracy     & Total         & Bandwidth & No. Episode  \\
                 &              & Time          &           & (Search Time)\\\hline
    MNIST-Acc         & 98.6\%  &  0.79\,s & 17\,MB  & 1000 (17 hrs.) \\\hline
    MNIST-Per         & 98.6\%  &  0.79\,s & 17\,MB  & 1000 (17 hrs.) \\\hline
    Fashion-Acc       & 90.6\%  &  1.67\,s & 50\,MB  & 2000 (32 hrs.) \\\hline
    Fashion-Per       & 90.4\%  &  0.72\,s & 22\,MB  & 2000 (32 hrs.) \\\hline
    CIFAR-Acc         & 84.6\%  &  8.0\,s  & 944\,MB & 1200 (60 hrs.) \\\hline
    CIFAR-Per         & 82.6\%  &  5.1\,s  & 582\,MB & 1200 (60 hrs.) \\\hline
  \end{tabular}
\end{table}

The predicted architectures laying on the Pareto frontier of the tested datasets 
are summarized in Table~\ref{tab:nass_performance}. Two types of
architectures are selected here. Architectures with a suffix of -Acc are the child 
networks that have better
accuracy but (relatively) worse performance, and -Per the reverse.
The insight here is that, for smaller (i.e., easier) datasets such as MNIST, the  
search is almost exhaustive, where the best architecture achieves highest 
prediction accuracy and cryptographic performance. Nevertheless, for more complex datasets, 
the differences become increasingly large, where distinctive trade-offs between
neural architectures emerge. We emphasize that depending on 
the application, all neural architectures on the Pareto frontier are legitimate 
candidates. However, as also shown in Fig.~\ref{fig:frontier}, in many cases, 
significant performance degradation only results in marginal accuracy improvement,
and vise versa. 

\subsection{Comparison to Existing Works}
\begin{table}
  \scriptsize
  \centering
  \renewcommand\arraystretch{1.35}
  \caption{Comparison Between Gazelle and Architecture from {\placeholder}}
  \label{tab:result}
  \begin{tabular}{ccc|ccc}
  \hline
    \multicolumn{3}{c|}{Gazelle} & \multicolumn{3}{c}{Best Searched by {\placeholder}}\\\hline
    Layer   & Dimension                  & Quant.& Layer & Dimension & Quant. \\\hline 
    CR & $(64\times 3\times 3)$     &  23     & CR & $(24\times 5\times 3)$ & $(8, 8)$\\
    CR & $(64\times 3\times 3)$     &  23     & CR & $(48\times 3\times 5)$ & $(6, 7)$\\
    PL & $(2\times 2)$              &   23    & PL & $(2\times 2)$ & $(8, 8)$\\
    CR & $(64\times 3\times 3)$     &  23     & CR & $(48\times 5\times 7)$ & $(7, 6)$\\
    CR & $(64\times 3\times 3)$     &  23     & CR & $(36\times 3\times 3)$ & $(6, 5)$ \\
    PL & $(2\times 2)$              &   23    & PL & $(2\times 2)$ & $(8, 8)$\\
    CR & $(64\times 3\times 3)$     &  23     & CR & $(24\times 7\times 1)$ & $(4, 6)$\\
    CR & $(64\times 3\times 3)$     &  23     & \\
    FC & $(1024\times 10)$     &  23     & FC & $(1024\times 10)$ & $(16, 16)$ \\\hline
    \multicolumn{3}{c|}{Accuracy: 81.6\%} & \multicolumn{3}{c}{Accuracy: 84.6\%}\\
    \hline
    \multicolumn{3}{c|}{Bandwidth: 1.815\,GBytes} & \multicolumn{3}{c}{Bandwidth: 977\,MB}\\
    \hline
    \multicolumn{3}{c|}{PAHE Time: 3.22\,s} & \multicolumn{3}{c}{PAHE Time: 1.62\,s}\\
    \multicolumn{3}{c|}{GC Time: 13.2\,s} & \multicolumn{3}{c}{GC Time: 6.38\,s}\\
    \multicolumn{3}{c|}{Total Time: 16.4\,s} & \multicolumn{3}{c}{Total Time: 8.0\,s}\\
    \hline
  \end{tabular}
\end{table}
We selected the CIFAR-Acc from Table~\ref{tab:nass_performance} to compare the 
{\placeholder} against the baseline
architecture proposed in~\cite{liu2017oblivious}. The architectures are summarized 
in Table~\ref{tab:result}. Here, CR depicts a
convolution layer plus a ReLU layer, and PL is an average pooling layer. 
Dimension indicates the filter
dimension, and the input dimension is $(3\times 32\times 32)$ in the CIFAR-10 dataset.
The important observation here is that, by using architectural search, 
we do not need to trade accuracy for performance. The generated neural architecture 
requires only 5 convolution layers rather than 6 (as used in the baseline architecture), 
while improving the prediction accuracy 
from 81.6\% to 84.6\%. The inference time and network bandwidth are reduced by 
2x and 1.9x, respectively. The reduction rate can be increased to more than 3x
when the same level of accuracy suffices, as demonstrated by the CIFAR-Per 
architecture in Table~\ref{tab:nass_performance}. Finally, we note that the very recent
work~\cite{riazi2019xonn} that achieves a prediction accuracy of 85\% requires more than 30 seconds
to carry out the inference, which translates to 4x time reduction when compared to CIFAR-Acc.


\section{Conclusion}\label{sec:conclusion}
In this work, {\placeholder} is proposed to optimize neural network
architectures used in secure inference schemes. Models of cryptographic
primitives are created to automatically generate computational and
communicational profiles. Rewards are generated based on the calculated
profiles and fed to a NAS optimizer to search in the architectural space of
convolutional neural networks. Experiments show that security-centric designs result in
better inference speed and bandwidth footprint compared to manually 
tuned neural architectures, while achieving better prediction accuracy.

\section*{Acknowledgment}
This work was partially supported by JSPS KAKENHI Grant No.~17H01713, 17J06952, Grant-in-aid for JSPS Fellow (DC1), National Science Foundation under Grant CNS-1822099, and Edgecortix Inc. 

\bibliographystyle{ecai}
\bibliography{cad,security,NAS}

\end{document}